\documentstyle[graphicx,aps,multicol]{revtex}

\draft
\begin{document}

\title{Energy shifted level densities in the rare earth region}

\author{M.~Guttormsen\footnote{E-mail: magne.guttormsen@fys.uio.no}, M.~Hjorth-Jensen, E.~Melby, J.~Rekstad, A.~Schiller and S.~Siem}
\address{Department of Physics, University of Oslo,
P.O.Box 1048 Blindern, N-0316 Oslo, Norway}

\maketitle

\begin{abstract}
The density of accessible levels at low spin in the ($^3$He,$\alpha \gamma$) reaction has been extracted for the $^{161,162}$Dy and $^{171,172}$Yb nuclei. The energy shift between the level densities of the even-odd and even-even isotopes is measured as a function of excitation energy. The results are compared with predictions from various semi-empirical models. The energy shift procedure works well for excitation energies between 3.5 and 7 MeV in the even-even nucleus, provided that a proper level density function is used. The experimental energy shift is close to the pairing gap parameter $\Delta$.
\end{abstract}


\begin{multicols}{2}

\section{Introduction}

The Fermi gas model for finite nuclei has been shown to account for the nuclear level density at high excitation energy \cite{1}. The model describes the nucleus as a gas of non-interacting fermions confined to the nuclear volume and neglects shell effects and pair correlations. 

The level density of odd-odd nuclei is relatively high already around the ground state due to the coupling of the two valence particles to the core. However, for odd-mass and even-even nuclei the level density is considerably lower due to the formation of Cooper pairs. A handful semi-empirical approaches has been suggested to describe this effect of pair correlations by a simple energy shift of the level density function.

In the conventional shifted Fermi gas model \cite{2,3,4} the excitation energy is shifted using the pairing energy parameter $\Delta$. The shifts are $\Delta$ and 2$\Delta$ for odd-mass and even-even nuclei, respectively, yielding approximately the level density found for the neighboring odd-odd system. This description turned out to be too rigid to reproduce the level densities at low and high excitation energies, simultaneously. A two-component level density formula with energy shifts was later introduced \cite{5,6}. Here, the first $\sim$ 10 MeV of excitation energy is described by a constant temperature formula, and at higher energies the shifted Fermi gas model is applied.

A simpler and rather well working version is the back-shifted\footnote{For light nuclei the ground state position has to be shifted to lower energies compared to the fit performed with the conventional shifted Fermi gas model; therefore the notation "back-shift".} Fermi gas model  \cite{7}, where the Fermi gas formula is used for all excitation energies. The model has only two parameters: the back-shifted energy and the level density parameter $a$, both being free parameters in order to fit the data.

There are several unclear points in using these approaches. The main questions concern the functional form of the level density and the justification of a shift of the excitation energy to describe the level densities of neighboring nuclei. The fact that the extracted shifts seldom coincide with the paring gap parameter $\Delta$ (or 2$\Delta$) indicates that one or both assumptions are not fulfilled.

In the vicinity of the ground band, levels can be counted reliably up to a certain excitation energy, typically 1.5 MeV in even rare earth nuclei. The level density can also be derived by the level spacing of neutron resonances at the neutron binding energy $B_n$. In between these energies few experimental results are available. In the A = 40 -- 60 mass region, fluctuation widths and charged particle emission spectra have been employed. Recently \cite{8}, the level density of $^{69}$As and $^{70}$Ge was measured using the ($^{12}$C,p) compound reaction. The data cover 5 -- 24 MeV of excitation energy, but eventual fine structures are smeared out due to the high average spin and broad spin distributions in such studies.

The Oslo group has extracted level densities from measured $\gamma$-ray spectra \cite{9,10,11,12}. The method, which is described in Ref.~\cite{13}, allows for the simultaneous extraction of the level density and the $\gamma$-strength function over a wide energy region. In these studies, the transferred spin in the ($^3$He,$\alpha$) reaction is approximately 2 -- 6 $\hbar$, and the nuclear system is believed to thermalize prior to $\gamma$-emission.

The subject of this work is to extract experimental energy shifts using the method of Ref.~\cite{13} and to investigate the quality of the energy shift procedure as function of excitation energy. Furthermore, it is interesting to compare the value of the energy shift parameter to the pairing gap parameter $\Delta$.

\section{Experimental method}

The experiments were carried out with 45 MeV $^3$He-projectiles at the Oslo Cyclotron Laboratory (OCL). The particle-$\gamma$ coincidences are measured with the CACTUS multidetector array \cite{14} using the ($^3$He,$\alpha \gamma$) reaction on $^{162,163}$Dy and $^{172,173}$Yb self-supporting targets. The charged ejectiles were detected with eight particle telescopes placed at an angle of 45$^{\circ}$ relative to the beam direction. An array of 28 NaI $\gamma$-ray detectors with a total efficiency of $\sim$15\% surrounded the target and particle detectors. 

The experimental extraction procedure and the assumptions made are described in Refs.~\cite{9,13} and references therein. The level density is deduced from $\gamma$-ray spectra recorded at a number of initial excitation energies $E$, determined by the measured $\alpha$-energy. These data are the basis for making the first generation (or primary) $\gamma$-ray matrix, which is factorized according to the Brink-Axel hypothesis \cite{15,16} as 
\begin{equation}
P(E,E_{\gamma}) \propto  \rho (E -E_{\gamma}) \sigma (E_{\gamma}),
\end{equation}
where the level density $\rho$ and the $\gamma$-energy dependent function $\sigma$ are determined by an iterative procedure.

To obtain the level density, the first trial function for $\rho$ is simply taken as a straight line and the corresponding $\sigma$ is determined by Eq.~(1). Then a $\chi^2$ minimum is calculated for each parameter, keeping the others fixed. This procedure is repeated about 50 times, until a global least square fit to the $\sim$ 1400 data points of $P$ is achieved.

It has been shown \cite{13} that if one solution for $\rho$ and $\sigma$ is found, functions of the form
\begin{equation}
\rho (E -E_{\gamma}) \rightarrow A \exp [\alpha(E-E_{\gamma})] \rho (E -E_{\gamma})
\end{equation}
and
\begin{equation}
\sigma (E_{\gamma}) \rightarrow B\exp (\alpha E_{\gamma})\sigma (E_{\gamma}),
\end{equation}
give exactly the same fit to the $P(E,E_{\gamma})$ matrix. The values of
$A$, $B$ and $\alpha$ can be determined by additional conditions. The $A$ and $\alpha$ parameters are used for absolute normalization of the level density $\rho$: They are adjusted to reproduce (i) the number of levels observed in the vicinity of the ground state and (ii) the neutron resonance spacing at the neutron binding energy $B_n$. Further details on the method, the iteration procedure and the simulation of errors are given in Ref. \cite{13}.

In the following we will not discuss the $\gamma$-energy dependent function $\sigma$, but concentrate on the level density $\rho$.

\section{Experimental and semi-empirical level densities} 

The experimental level densities for the $^{161,162}$Dy and $^{171,172}$Yb nuclei are shown as data points in Figs.~1 and 2. In the extraction technique, we exclude data with $\gamma$-energies below 1 MeV due to methodical problems in the first generation spectra. Therefore, the level density is generally determined only up to $E = B_n - 1$ MeV. Recently \cite{12}, thermodynamical aspects have been discussed from the very same level density curves as shown here. The figures include the level densities (solid drawn lines) obtained from counting known levels [18]. These known densities are seen to agree with our experimental data points. The level densities from counting are reliable up to an excitation energy of $E \sim$ 0.8 MeV and 1.7 MeV for the odd- and even-mass cases, respectively.  Above these energies, more than 100 levels are present per MeV and it is experimentally difficult to resolve all levels. Therefore, the density of known discrete levels drops at higher excitation energies.

The level densities for the $^{161}$Dy and $^{171}$Yb isotopes are about five times higher than for the neighboring $^{162}$Dy and $^{172}$Yb isotopes. The latter isotopes seem to exhibit the same slope at high excitation energy. However, the presence of bumps modifies this simple picture, in particular at low excitation energies.

The energy region up to $\sim 5 - 10$ MeV has been described by the constant temperature formula [5,6] given by
\begin{equation} 
\rho=C \exp (E/T),
\end{equation}
where the normalization factor $C$ and the temperature $T$ are constants. Also level densities based on the Fermi gas model are frequently adopted in this energy region~\cite{5,6,7} 
\begin{equation}
\rho=\frac{\exp [2\sqrt{aU}]}{12\sqrt{2}a^{1/4}U^{5/4}\sigma},
\end{equation}
where $\sigma$ is the spin cut-off parameter and $U$ is the back-shifted energy. As examples of such approaches Figs.~1 and 2 also include level densities from Gilbert and Cameron [5] (dashed curves) and from von Egidy et al.~[6] (dash-dotted curves). Full details on the formulae, parameterizations and choice of parameters are given in Refs.~[5,6].

The level densities of Gilbert and Cameron are described by Eq.~(4) in the excitation region below $\sim 5$ MeV, and at higher energies they use Eq.~(5). The description (dashed curves) is rather poor, except for $^{171}$Yb. Gilbert and Cameron give temperatures that are lower in the even-even systems, contrary to the tendency of our data\footnote{The temperature $T$ can be extracted from the inverse of the slope of the logarithm of the data points, i.e., $1/T=d\ln \rho /dE$.}. It seems that these authors have, for the even-even nuclei, anchored their constant temperature level density curves to the ground state band, rather than to levels at $\sim 1.7$ MeV, where the two quasiparticle regime appears. Also scarce data at the time their compilation was made (1965) could be the reason for the poor agreement. Even so, we think the two-component level density is a reasonable approach. In the first MeV of excitation energy, nucleon pairs (Cooper pairs) are broken and thus prevent the temperature to rise as fast as predicted by the Fermi gas formula. This mechanism is discussed in Refs.~[11,12], and references therein. For excitation energies around and above the neutron binding energy, the Fermi gas conditions are probably fulfilled. Here, the pairing correlations are quenched and a high density of single particle levels is present. 
 
von Egidy et al.~[6] have tested both the constant temperature and the back-shifted Fermi gas formulae in this region. They find that both approaches give similar $\chi ^2$-fits to experimental data. The suggested temperatures are close to 0.6 MeV for all four nuclei, almost 0.1 MeV higher than our data indicate. In Figs.~1 and 2 we show only the Fermi gas results (dash-dotted curves). Within this model both the level density parameter $a$ and the correction to the back-shift $C_1$ ($U = E - \Delta - C_1$) are based on global parameterizations as function of the mass number $A$. With this restriction, one may say that the level density curves describe our data points rather well. However, a clear shortcoming is that these expressions increase too slowly as function of excitation energy. The experimental level densities, in particular for the Yb isotopes, show a functional form closer to the constant temperature formula\footnote{Gilbert and Cameron obtain a poor constant temperature fit for $^{172}$Yb, probably because they try to fit the level density of the ground band.}.

Global fits for all mass numbers can give deviations of up to a factor of 10 from known average neutron resonance spacings [6]. Of course, better local fits to experimental data could be achieved, both with Eq.~(4) and/or Eq.~(5). However, a common approach is difficult to construct since all four nuclei exhibit different functional forms. These variations are probably connected with details in the quenching of the pair correlations in the individual nuclei. 

In the following, we will not propose any new semi-empirical formulae, but rather focus on the feature that the level densities for the neighboring isotopes exhibit similar shapes as function of excitation energy.

\section{Comparison between experimental and semi-empirical energy shifts}

It is commonly believed that neighboring odd-odd, odd-even, even-odd and even-even isotopes reveal the same level density if a proper shift is applied to the excitation energy. This is indicated in Fig.~3, where the energies are compared to the ground state of the odd-odd nucleus. The even-even nucleus has its ground state pushed down by $\Delta_p + \Delta_n$ relative to the odd-odd system due to pairing interactions. The odd-mass isotopes represent cases in between these limits. With the present experimental data, we have the opportunity to test how well the energy-shift procedure works for the level densities of even-odd (eo) and even-even (ee) systems.  

Neglecting collective excitations and two-body forces, various models can describe the level density of the odd-odd (oo) nucleus rather successfully, e.g. the Fermi gas model. The level density of the other neighboring nuclei can then be estimated by
\begin{eqnarray}
\rho _{\rm oe}(E) = \rho _{\rm oo}(E-\Delta_n),\\
\rho _{\rm eo}(E) = \rho _{\rm oo}(E-\Delta_p),\\
\rho _{\rm ee}(E) = \rho _{\rm oo}(E-\Delta_n-\Delta_p),
\end{eqnarray}
where $E$ is the excitation energy. 

The pairing gap parameters $\Delta_p$ and $\Delta_n$ can be determined from empirical masses of a sequence of isotones or isotopes where \cite{17}
\begin{eqnarray}
\Delta_p&=&\frac{1}{4}|S_p(N,Z+1)-2S_p(N,Z)+S_p(N,Z-1)|,\\
\Delta_n&=&\frac{1}{4}|S_n(N+1,Z)-2S_n(N,Z)+S_n(N-1,Z)|,
\end{eqnarray}
and $S_p$ and $S_n$ are proton and neutron separation energies \cite{18}, respectively. The pairing gap parameter can alternatively be calculated by the empirical formula \cite{17}
\begin{equation}
\Delta = 12A^{-1/2} {\rm MeV},
\end{equation}
which is valid for both neutrons and protons. 

Equations (9) and (10) depend on the proton ($Z$) and neutron ($N$) numbers and should in principle give the best estimate. However, Eq.~(11) gives a smooth function which neglects local shell effects, and is probably more correct if $\Delta$ is interpreted as a pure pairing parameter.

From the extracted level densities for the $^{161,162}$Dy and $^{171,172}$Yb nuclei, we can investigate the energy shift necessary to apply in order to simulate the level density in neighboring even-odd and even-even isotopes. The energy shift $\delta(E)$ is defined as the necessary shift of the even-odd nucleus level density in order to describe the level density in the neighboring even-even nucleus
\begin{equation}
\rho_{\rm ee}(E) = \rho_{\rm eo}(E-\delta(E)).
\end{equation}
In Fig.~4 the resulting $\delta(E)$ curves are plotted as function of the excitation energy $E$ measured in the even-even nucleus. In the excitation energy region between 3.5 and 7 MeV the energy shift is rather constant giving $\delta =$ 1.13(7) and 0.84(10) MeV for $^{161,162}$Dy and $^{171,172}$Yb, respectively. 

The corresponding $\delta$-values should also compare with the effective pairing gap parameters of
\begin{equation}
\Delta ^{\rm eff}(^{162}{\rm Dy})=\Delta_p(^{162}{\rm Dy})+\Delta_n(^{162}{\rm Dy})-\Delta_p(^{161}{\rm Dy})
\end{equation}
and
\begin{equation}
\Delta ^{\rm eff}(^{172}{\rm Yb})=\Delta_p(^{172}{\rm Yb})+\Delta_n(^{172}{\rm Yb})-\Delta_p(^{171}{\rm Yb}),
\end{equation}
where we apply Eqs.~(9) and (10) for the pairing gap parameters. In Table 1 these values (and the pairing gap parameters calculated from Eq.~(11)) are compared to the experimental $\delta$-values. These values coincide rather well within less than 0.2 MeV, and Fig.~4 shows that $\Delta^{\rm eff}$ is in good agreement with the observed energy shifts in the 3.5 -- 7 MeV excitation region. 
The agreement is not that impressive when comparing experiment with energy shifts obtained from Eq.~(11) and with semi-empirical level density functions. The shifts from Gilbert and Cameron [5] (dashed curves in Fig.~4) deviate strongly from the experimental data, as also indicated from the $\delta$-values\footnote{Since these authors apply  different temperatures for even-odd and even-even nuclei (see Figs.~1 and 2), the actual shifts are only approximately given by the $\delta$-values of Table 1.} of Table 1. The shifts from von Egidy et al.~[6] (dash-dotted curves) are determined by the experimental pairing gap $\Delta$ and the slow varying back-shift correction $C_1$. Therefore, these shifts coincide almost excactly with the $\Delta ^{\rm eff}$ values. The small deviations seen in Fig.~4 are due to the 0.1 MeV$^{-1}$ increase in the level density parameters for the even-even systems.

Both the two-component formula of Gilbert and Cameron~\cite{5} as well as the Fermi gas formula of Ref.~\cite{7} give $\delta$-values deviating with about 0.3 -- 0.5 MeV (see Table 1). This is probably due to the free adjustment of $\delta$ and other parameters, and indeed the shifts have been associated with large uncertainties by these authors. The role of $\delta$ in this type of approaches is not a pure energy shift, but may also include a compensation for the unphysical form of the adopted analytical level density function. The same conclusion is evident from the compilations of Refs.~\cite{5,6,7}, where the extracted energy shifts scatter typically within $\pm$ 0.5 MeV in this mass region.

\section{Conclusion}

The shifting of excitation energy in order to simulate the level density of neighboring isotopes works well using realistic level density functions. The level densities follow each other rather close as function of excitation energy in the 3.5 -- 7 MeV region, and the energy shifts coincide within 0.2 MeV with the pairing gap parameter $\Delta$. Of the approaches studied here, only the formalism of von Egidy et al.~has this feature included. Below 3.5 MeV of excitation energy, nuclear structures assigned the various nuclei prevent the use of a simple energy shift procedure. In particular, the even-even isotopes reveal bumps in the level density function due to the breaking of Cooper pairs. 

The application of analytical level density functions may give energy shifts that deviate significantly from the experimental pairing gap parameters. In this work we have demonstrated that the functional form and/or the parameters used are not fully appropriate for the description of level densities below 7 MeV of excitation energy. 

Probably, no simple level density formula can describe simultaneously the four nuclei investigated here. The Yb isotopes exhibit a constant temperature-like behavior, while the Dy isotopes are closer to the back-shifted Fermi gas prediction. Nevertheless, we find that the parameters used for the semi-empirical formulae should undergo a revision. Here, all new low-lying levels should be included together with recent information on resonance level spacings. This effort, combined with a refined two-component formula, like the one of Gilbert and Cameron, could probably give better analytical formulae for future use. The analytical expressions should have the ability to give a constant energy shift between the level densities of neighboring isotopes, as observed in this work for the $^{161,162}$Dy and $^{171,172}$Yb nuclei.

The authors are grateful to E.A.~Olsen and J.~Wikne for providing the excellent experimental conditions. We wish to acknowledge the support from the Norwegian Research Council (NFR).

\end{multicols}

\begin{table}
Table 1: Energy shift $\delta$ extracted between the even-odd and even-even isotopes$^a$.
\begin{tabular}{l|c|c}
Parameter $\delta$ or $\Delta$ (MeV)                & $^{162}$Dy  &$^{172}$Yb\\ \hline
$\delta$ from present data                          & 1.13(7)     &  0.84(10)\\
$\Delta^{\rm eff}$ from separation energies, Eqs.~(13,14) & 1.05   &  0.93\\
$\Delta$ from empirical formula, Eq.~(11)           & 0.94        &  0.91\\
$\delta$ from back-shifted Fermi gas [7]            & 0.88(50)    &  1.15(50)\\ 
$\delta$ from two-component level density [5]:      &             &   \\
Energies below $\sim$ 5 MeV, Eq.~(4)                & 0.81(30)$^b$&  1.30(30)\\
Energies above $\sim$ 5 MeV, Eq.~(5)                & 0.70(20)    &  0.69(20)\\
\end{tabular}
$^a$ The energy shifts from semi-empirical level densitiy formulae are constants based on parameters used in Refs.~[5-7]. The energy shifts of von Egidy et al.~[6] are not listed since they coincide with $\Delta^{\rm eff}$.\\
$^b$ The shift is calculated from the $^{160,161}$Dy parameter sets.
\end{table}

\begin{figure}
\includegraphics[totalheight=17.5cm,angle=0,bb=0 80 350 730]{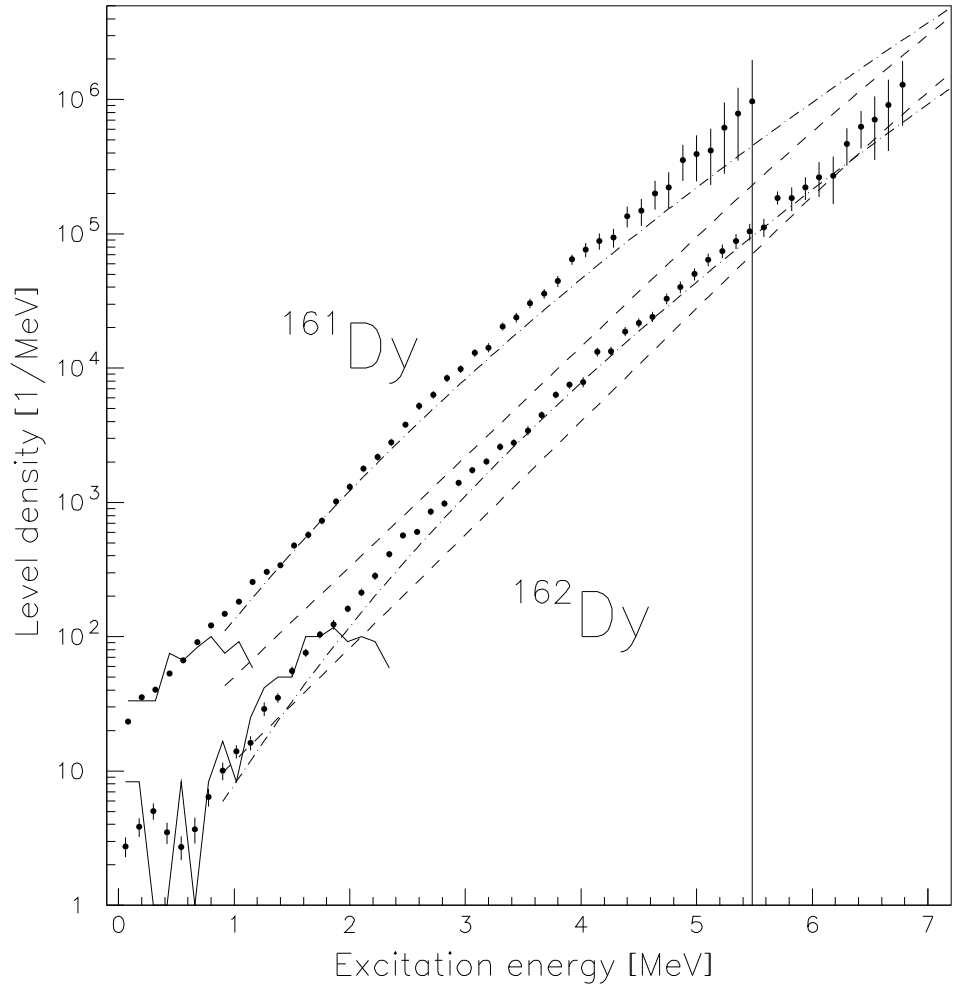}
\caption{Observed level densities for $^{161,162}$Dy as functions of excitation energy. The experimental data points are compared to the density of known levels at low excitation energy (solid lines).  The figure also includes the semi-empirical level density formulae of Gilbert and Cameron [5] (dashed curves) and von Egidy et al.~[6] (dash-dotted curves). Upper and lower points/curves are for $^{161}$Dy and $^{162}$Dy, respectively. Since Gilbert and Cameron give no parameters for $^{162}$Dy, we use the $^{160}$Dy parameter set.} 
\end{figure}

\begin{figure}
\includegraphics[totalheight=17.5cm,angle=0,bb=0 80 350 730]{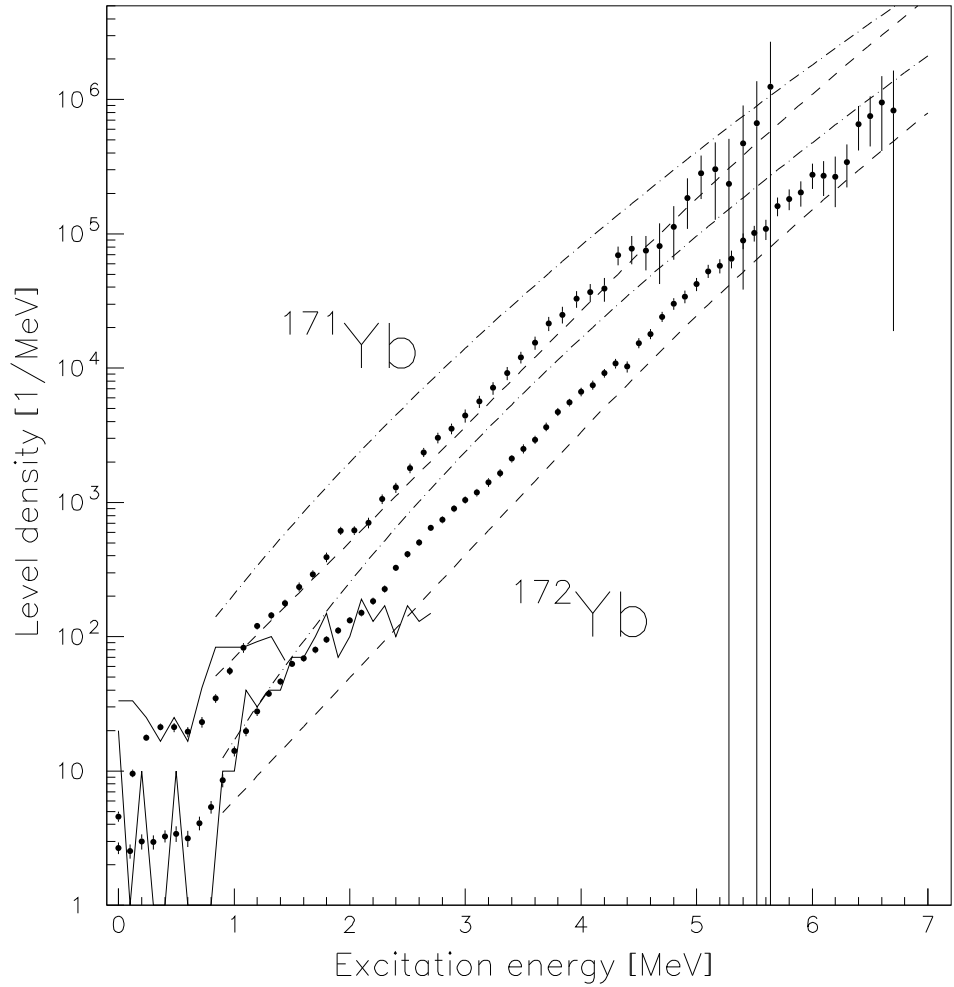}
\caption{Observed level densities for $^{171,172}$Yb as functions of excitation energy. The experimental data points are compared to the density of known levels at low excitation energy (solid lines) and the semi-empirical level density formulae of Gilbert and Cameron [5] (dashed curves) and von Egidy et al.~[6] (dash-dotted curves). Upper and lower points/curves are for $^{171}$Yb and $^{172}$Yb, respectively.}
\end{figure}

\begin{figure}
\includegraphics[totalheight=17.5cm,angle=0,bb=0 80 350 730]{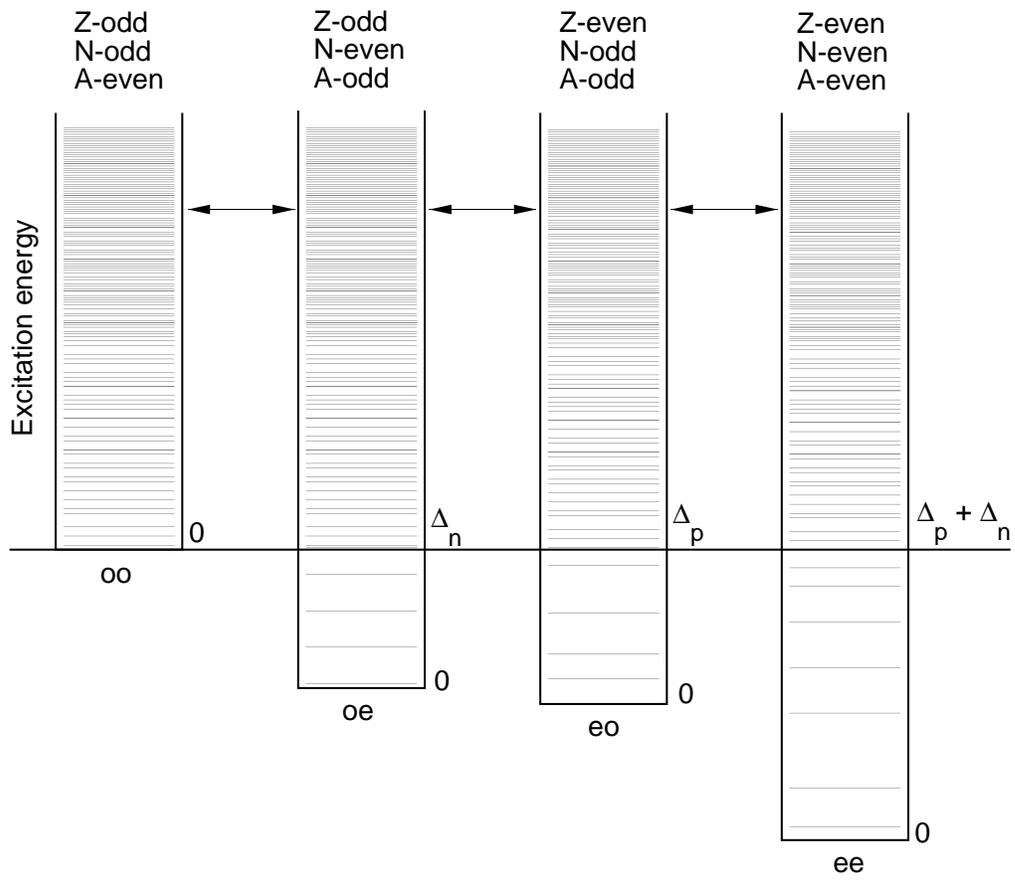}
\caption{Illustration of how the level density for various nuclei (oo, oe, eo, and ee) can be estimated by proper energy shifts.} 
\end{figure}

\begin{figure}
\includegraphics[totalheight=17.5cm,angle=0,bb=0 80 350 730]{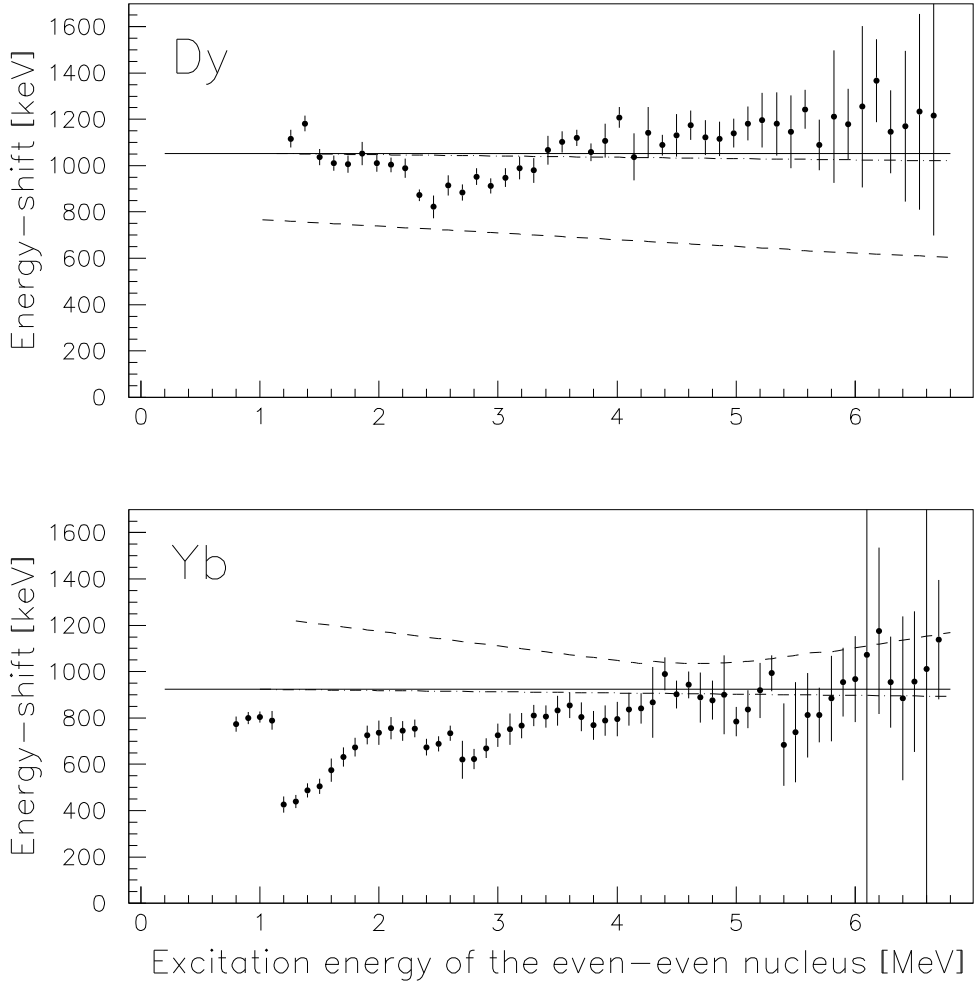}
\caption{The observed energy shift $\delta$ between $^{161}$Dy and $^{162}$Dy (upper part) and $^{171}$Yb and $^{172}$Yb (lower part). The pairing gap parameters $\Delta^{\rm eff}$, evaluated from Eqs.~(13) and (14), are displayed as solid lines for comparison. The energy shifts obtained from Gilbert and Cameron [5] (dashed curves) and von Egidy et al.~[6] (dash-dotted curves) are also displayed.}
\end{figure}

\end{document}